\documentclass[preprint,groupedaddress,floatfix,%
nofootinbib]{revtex4}

\usepackage{dcolumn}
\usepackage{graphicx}
\usepackage{epsfig}
\usepackage{hyperref}
\usepackage{graphicx}
\usepackage{dcolumn}
\usepackage{amssymb,eucal}
\usepackage{longtable}
\usepackage{amsmath,amssymb,bm}
\usepackage{mathrsfs}

\newcommand{\mathd}{\mathrm{d}}
\newcommand{\mathi}{\mathrm{i}}
\newcommand{\mathe}{\mathrm{e}}

\newcommand{\nn}{\nonumber}

\begin{document}

\title{A note on electromagnetic edge modes}

\author{Fen Zuo\footnote{Email: \textsf{zuofen@hust.edu.cn}}}
\affiliation{School of Physics, Huazhong University of Science and Technology, Wuhan 430074, China}

\begin{abstract}
We give an intuitive identification for the electromagnetic edge modes as virtual spinon excitations in quantum spin liquids.
Depending on the statistics, these edge modes could be effectively described by the $\beta \gamma$ or $bc$ conformal field theory. As an example, we show how such a description may reconcile the discrepancy on the logarithmic coefficient of the entanglement entropy on a sphere. Also we give some comments on the possibility of a topological term in the entanglement entropy.

\end{abstract}
 \maketitle
\section{Entanglement entropy for electromagnetic field}
The evaluation of the entanglement entropy~(EE) is usually complicated. Only in special cases, the calculation could be analytically performed. For a general conformal field theory~(CFT) with spherical entangling surface, one could employ the conformal transformations to convert EE into thermal entropy in the transformed spacetime, evaluated at a fixed temperature $T=1/2\pi R$~\cite{Casini:2011kv}. With different choices of the transformation, the resulting spacetime could be the static patch of de Sitter space, or the open Einstein universe $R\times H^3$. Here we focus on the entanglement of electromagnetic fields in $3+1 D$ Minkowski spacetime, with the entanglement surface being a sphere with radius $R$. The global thermal analyses in de Siter space dictates that EE contains a universal logarithmic term as follows
\begin{equation}
S_{\rm {CFT}}\sim -4a \log \frac{R}{\epsilon},\label{eq.anomaly}
\end{equation}
with $\epsilon$ is the UV regulator. $a$ is the $a$-type central charge, which is related to the coefficient of the Euler density in the trace anomaly. When the CFT has a gravity dual description, such a universal term can be obtained from the Ryu-Takayanaki formula~\cite{Ryu:2006bv,Ryu:2006ef,Nishioka:2009un}. For the electromagnetic field, $a=62/360$~\cite{Birrell-Davies1982}, resulting
\begin{equation}
S_{\rm{ YM}}\sim -\frac{62}{90} \log \frac{R}{\epsilon}.\label{eq.YM1}
\end{equation}
However, a direct local calculation of the thermal entropy in de Sitter space and the open Einstein universe shows~\cite{Dowker:2010bu,Eling:2013aqa}
\begin{equation}
\tilde S_{\rm{ YM}}\sim  -\frac{32}{90} \log \frac{R}{\epsilon}.\label{eq.YM2}
\end{equation}
The coefficient differs from that in (\ref{eq.YM1}) by $-1/3$. The controversy between (\ref{eq.YM1}) and (\ref{eq.YM2}) is later attributed to ignoration of the edge modes on the entangling surface in the latter~\cite{Donnelly:2014fua,Huang:2014pfa,Donnelly:2015hxa}, since the local calculation in~\cite{Dowker:2010bu} counts only the bulk modes. According to the classification in~\cite{Casini:2013rba}, this corresponds to the local operator algebra with a trivial center. In~\cite{Donnelly:2014fua,Donnelly:2015hxa} it is shown that if instead the pure electric center is considered, the edge modes contribute as a ghost scalar and (\ref{eq.YM1}) is recovered.
It is proposed as the Fadeev-Popov ghosts to the gauge-fixing procedure on the edge~\cite{Huang:2014pfa}. The result (\ref{eq.YM2}) is confirmed recently through a direct calculation of EE employing the duality between the Maxwell field and two massless scalar fields, with the zero modes of the latter subtracted~\cite{Casini:2015dsg}. Such a calculation also shows clearly that it is completely due to the bulk gauge contribution. With these results, the discrepancy is more or less clarified. However, there are still some doubts on the edge modes, as emphasized in~\cite{Casini:2015dsg}. One question is, are the ghosts physical excitations, or simply re-parametrization fields? Such a question is recently considered in a similar setting, namely boundary CFT reconstruction of Wilson line operators across the horizon in an Anti de-Sitter Schwarzschild background~\cite{Harlow:2015lma}. Another question is, what will be the corresponding result in the magnetic center?


\section{Spinons as edge modes}
We try to answer these questions from a different point of view, treating Maxwell theory as emerging from some underlying spin system, when the system is in the so-called quantum spin liquid phase~\cite{Anderson:1987}.
The spin system possesses $SU(2)$ symmetry, which breaks down to $U(1)$ due to the dynamics. Emergence of $U(1)$ gauge dynamics could be seen in various ways, which usually employ the ``slave-particle'', or ``parton'', construction. Typical ways involve parameterizing the spin with Abrikosov fermions or Schwinger bosons~(see, e.g., \cite{Wen2004,Lee:2006zzc,Fradkin2013} and references therein). In the quantum rotor model~\cite{Wen2004,Hermele2004,Levin:2005vf,Savary2011}, the gauge structure is most clearly seen.
The underlying mechanism for emergence of the massless photons is proposed to be string-net condensation~\cite{Levin:2004mi}.

Sting-net condensation also provides a nice physical picture for the calculation of entanglement entropy. It could be viewed from two different pictures: the string-net picture and the closed string picture~\cite{Wen2004}. Here for the $U(1)$ case we focus on the closed string picture, as recently considered in \cite{Pretko:2015zva}. The closed strings are simply the electric flux lines of the emergent gauge fields. In some parameter regime of the spin system, these strings have an almost vanishing tension and therefore proliferate rapidly. The only low-energy excitation is the massless photon, arising from the fluctuation of the closed strings. Only at very high energy, the open strings can be excited, with the endpoints carrying electric charges. So these real excitations will not be present in our calculation of the entanglement entropy of the electromagnetic field. However, when we decompose the full system into two subsystems, some of the close strings will be cut into open ones. Due to the condensation, the string part of the open strings are indistinguishable from the local closed ones in a subsystem. The endpoints act as effective charge excitations on the entangling surface. In particular, with such virtual charge excitations the total Hilbert space is factorizable, since the original microscopic system is local. So the calculation of the entanglement entropy is straightforward. As a result, the entanglement entropy will be clearly separated into two parts, the bulk gauge contributions and the edge particle contributions. Additional arguments for such a separation are given in~\cite{Pretko:2015zva}, based on the Bisognano-Wichmann theorem~\cite{Bisognano:1976za}. 
The bulk gauge part is cleanly extracted from the local thermal analyses~\cite{Dowker:2010bu} and the duality to the scalars~\cite{Casini:2015dsg}. We now try to extract the edge contribution, assuming that they are virtual excitations on the entangling surface. In quantum spin liquids typical particle excitations are spinons and visons, in addition to the emergent photons~\cite{Wen2004,Hermele2004,Levin:2005vf,Lee:2006zzc}. The spinons are excited when the closed electric strings are broken, thus representing the edge modes of our concern. Similar setup has been implicitly used in the derivation~\cite{Levin2006,Hung:2015fla} of the topological entanglement entropy~\cite{Kitaev:2005dm,Levin2006} in the toric code model~\cite{Kitaev:1997wr} and discrete string-net models~\cite{Levin:2004mi}.
  .
%



\subsection{Spinon dynamics: a review}
In the following we give a short review of the properties of the spinons, taking the notation in~\cite{Savary2011,Hao:2014rpa}~(with slight modifications). The spin system is defined on a bi-partite lattice, with two sublattice I and II. The sublattice sites could be distinguished by $\eta_{r}=\pm1$. The $L$ vectors starting from sites in sublattice I are denoted as $\hat \mu$, with $1\le\mu\le L$. Spins are defined on the links, with the following ``slave-particle'' parametrization~\cite{Savary2011}:
\begin{equation}
S_{rr'}^+=\psi_r^\dagger s^+_{rr'} \psi_{r'},~~~~S_{rr'}^-=\psi_{r'}^\dagger s^-_{rr'} \psi_{r},~~~~S^z_{rr'}=s^z_{rr'}.
\end{equation}
$\psi^\dagger$ and $\psi$ are the raising and lowering operators for the spinon number operator $Q$, defined as
\begin{equation}
\psi^\dagger \equiv \mathe ^{\mathi \varphi},~~~~\psi \equiv \mathe ^{-\mathi \varphi}.
\end{equation}
So $\varphi$ is the canonical conjugate variable to $Q$, with the commutation relation $[\varphi,Q]=\mathi$.
The gauge fields are then identified as
\begin{equation}
s_{rr'}^{\pm}\equiv\mathe^{\pm\mathi \phi_{rr'}},~~~~A_{rr'}=\eta_r \phi_{rr'}, ~~~~~E_{rr'}=\eta_r s^z_{rr'},
\end{equation}
together with the Gauss's law
\begin{equation}
Q_r=\sum _{r'} E_{rr'}.
\end{equation}
Here the summation is over all the links starting from $r$. Notice that $A_{rr'}$ and $E_{rr'}$ defined in this way are oriented. The above mapping shows manifestly the relation to the gauge theory. In particular, the $U(1)$ symmetry is the physical rotation symmetry around the $z$ axis in the spin space.

Due to the spin interaction from nearest neighbor links, the spinons can only hop within the same sublattice. Therefore we get two copies of identical spinons. Acting with $S_{rr'}^+$ creates a spinon on one sublattice site and an anti-spinon on the other. The dynamics of spinons is inherited from the spin dynamics, and takes the form:
\begin{equation}
H_s=-\sum _{r\in I}\sum_{\mu<\nu}[\psi_r^\dagger \psi_{r+\hat \mu -\hat \nu}+\psi_{r+\hat \mu}^\dagger \psi_{r+\hat \nu}+\mbox{H.c.}],\label{eq.hop}
\end{equation}
where `H.c.' stands for the hermitian conjugate terms.

\subsection{CFT description on the entangling surface}
In the continuum limit on a $2D$ entangling surface, one expects $\varphi$ to be described by a free massless scalar. Bosonization then tells us that $\psi/\psi^\dagger$ correspond to the $bc$ system with equal weights $h_b=h_c=1/2$~\cite{Polchinski-1998}.
In particular, since the $U(1)$ symmetry is the rotation symmetry around the $z$ axis in the spin space, spinons with different charges are described by different spin components of a $2D$ field. Focusing on the sublattice I, we thus have a spinon field $\psi^1$ with spin $1/2$, and an anti-spinnon field $\tilde \psi^1$ with spin $-1/2$. Their dynamics, inherited from the hopping term (\ref{eq.hop}), could be implemented through the following action
\begin{equation}
S_1=\frac{1}{4\pi}\int \mathd ^2 z (\psi^1 \bar \partial \psi^1 + \tilde \psi^1 \partial  \tilde \psi^1 ),
\end{equation}
where $\bar \partial \equiv \partial _{\bar z}$, $\partial\equiv \partial_z$ with $z$ and $\bar z$ the complex coordinates. On the other sublattice we have simply another copy. One could further combine them
\begin{eqnarray}
&&\psi=(\psi^1+\mathi \psi^2)/\sqrt{2},~~~ \bar \psi =(\psi^1-\mathi \psi^2)/\sqrt{2},\nn\\
&&\tilde \psi=(\tilde \psi^1+\mathi \tilde \psi^2)/\sqrt{2},~~~\bar {\tilde \psi}=(\tilde \psi^1-\mathi \tilde \psi^2)/\sqrt{2}.
\end{eqnarray}
Then the full action is
\begin{equation}
S=\frac{1}{2\pi}\int \mathd ^2 z (\psi \bar \partial  \bar \psi + \tilde \psi \partial \bar {\tilde \psi} ).
\end{equation}
So we indeed get the $bc$ system as expected. The conformal weights of $\psi$ and $\bar \psi$ are both $(1/2,0)$, giving the same spin $1/2$.

How about the statistics between these fields? The above mentioned bosonization procedure leads naturally to fermionic spinons~\cite{Polchinski-1998}. But at the lattice level, these fields could be just local bosonic operators~\cite{Wen2004,Hermele2004,Levin:2005vf,Savary2011}. Such a choice seems to be more natural since the spins on the links are simply bosonic. With some kind of ``twist'' construction~\cite{Levin:2004mi,Levin:2005vf}, the spinons could indeed be fermionic, according to the lattice definition of statistics through the hopping operator algebra~\cite{Levin:2003ws}. When restricted on the closed-loop sector, both the untwisted and twisted constructions give rise to the same low energy $U(1)$ gauge theory. And these two are two only possibilities in $3+1D$ and higher dimensions~\cite{Levin:2005vf}. The statistics of the spinons in $U(1)$ gauge theory is analyzed in detail recently~\cite{Wang2015}. Therefore for the untwisted theory, we have a $\beta \gamma$ ghost system on the entangling surface, with central charge $c_b=-1$. And in the twisted case, we have a $bc$ system with $c_f=1$. Such a generalized ghost system is discussed in detail in~\cite{CFT} and \cite{Polchinski-1998}. Entanglement entropy in some ghost CFTs is investigated recently in \cite{Narayan:2016xwq}.


One may find that the above description is very similar to the ``exclusive bosons'' framework proposed in \cite{Hao:2014rpa}. It will be interesting to further explore the exact relation between the two.


\subsection{Entanglement entropy and logarithmic coefficient}
Now we can calculate the edge contribution to the entanglement entropy, with focus on the logarithmic coefficient. According to the standard procedure~\cite{Birrell-Davies1982}, this is directly related to the central charge of the edge theory.
For the untwisted gauge theory, we have bosonic spinons with $c_b=-1$, giving the logarithmic term
\begin{equation}
S_b \sim -\frac{1}{3}\log \frac{R}{\epsilon}.\label{eq.Bspinon}
\end{equation}
Adding this to (\ref{eq.YM2}), one recovers (\ref{eq.YM1}) as in~\cite{Donnelly:2014fua,Donnelly:2015hxa,Huang:2014pfa}. This is to be expected, since starting from the bosonic gauge links one naturally gets bosonic spinons.
For the twisted theory, the edge modes are composed of fermionic spinons. The central charge $c_f=1$ dictates that
\begin{equation}
S_f \sim \frac{1}{3}\log \frac{R}{\epsilon}.\label{eq.Fspinon}
\end{equation}
It seems that it is impossible to recover such a term starting from the $U(1)$ gauge theory. In other words, the twist structure is lost in the low energy theory.

From the view point of string-net condensation, the above two results exhaust all the possibilities for the emergent gauge theory in $3+1 D$~\cite{Levin:2004mi}. In particular, the trivial and magnetic centers proposed in~\cite{Casini:2013rba} are not sufficient to characterize the edge modes for a string-net condensation. Recent discussion on the algebra centers in gauge theory can be found in \cite{Ma:2015xes}. Naively, one may suspect that the magnetic center will correspond to the edge modes consisting of visons, or magnetic monopoles. However, the visons alone can not specify all the boundary possibilities of the closed electric loops.

In the end of this section, we would like to add some comments on the possible topological term in the entanglement entropy. Recently in~\cite{Pretko:2015zva}~(see also~\cite{Radicevic:2015sza}), it is suggested that the charge neutral condition on the entangling surface leads to a topological term, which takes the following form when the radius $R$ is large
\begin{equation}
S_{\rm{ top}}\sim -\log \frac{R}{\epsilon}.\label{eq.top}
\end{equation}
If this is true, it immediately violates the previous consistence between the global anomaly analysis (\ref{eq.YM1}) and the local thermal results, (\ref{eq.YM2}) plus (\ref{eq.Bspinon}). However, from our CFT description it is not difficult to show that such a term will not be present. Since the spinons and anti-spinons are described by the holomorphic and anti-holomorphic parts of the CFT, they have exactly the same thermal distribution on the entangling surface at finite temperature. So their charges compensate exactly, leading to a neutral system. Alternatively, one can say the net charge should be zero since the chemical potential is zero. This is in contrast to the non-Abelian case studied in \cite{Zuo:2015mxk}, where the colorless constraint on the entangling surface indeed gives rise to a topological term. One may naively take the $N\to 1$ limit of the $U(N)$ theory there, and find the matrix integral becomes trivial.

\section{Discussion}
In the paper we have tried to understand the entanglement structure of electromagnetic field from the emergent point of view. We propose that the edge modes on the entangling surface could be identified as the virtual spinon excitations. Such an understanding is very similar to the recent proposal in~\cite{Harlow:2015lma}. In $3+1D$ theory, these edge modes are proposed to be described by some specific CFTs. This seems to give a natural explanation of the inconsistency on logarithmic coefficients of the entanglement entropy on a sphere. It will be interesting to further check the corresponding results with other entangling surfaces.


\section*{Acknowledgments}
The author benefits from the workshop ``Entanglement at Fudan 2015---where quantum information, condensed matter and gravity meet'', held at Fudan University, Shanghai, China. He would like to thank Ling-Yan Hung for the invitation.
The work is partially supported by the National Natural Science Foundation of China under Grant No. 11405065 and No. 11445001.

%
\end{document}